\documentstyle[prl,aps]{revtex}

\begin{document}
\input epsf
\draft
\tighten

\def\bk {{ \mathbf{k} }}
\def\bkappa   { \mbox{\boldmath ${\kappa}$}   }


\preprint{mpi-pks/9712006
STPHY 28/96  HEPHY-PUB 677/97
(to be submitted to Phys Rev E)
}

\title{Spatial correlations of singularity strengths in multifractal 
branching processes}

\author{
Martin Greiner$^1$, 
J\"urgen Schmiegel$^2$, 
Felix Eickemeyer$^2$,
Peter Lipa$^3$,
and Hans C.\ Eggers$^4$
}

\address{$^1$Max-Planck-Institut f\"ur Physik komplexer Systeme, 
             N\"othnitzer Str.\ 38, D--01187 Dresden, Germany}
\address{$^2$Institut f\"ur Theoretische Physik, Technische Universit\"at,
             D--01062 Dresden, Germany}
\address{$^3$Institut f\"ur Hochenergiephysik,
             \"Osterreichische Akademie der Wissenschaften, 
             Nikolsdorfergasse 18, A--1050 Vienna, Austria}
\address{$^4$Department of Physics, University of Stellenbosch,
             7600 Stellenbosch, South Africa}

\date{12.12.1997  chao-dyn/9804028}

\maketitle

\begin{abstract}
The $n$-point statistics of singularity strength variables for 
multiplicative branching processes is calculated from an 
analytic expression of the corresponding multivariate generating 
function. The key ingredient is a branching generating function 
which can be understood as a natural generalisation of the
multifractal mass exponents. Various random multiplicative cascade 
processes pertaining to fully developed turbulence are discussed.
\end{abstract}




\section{Introduction}                      \label{intrd}

Many complex processes occurring in nature are believed to be
organized in a selfsimilar way: turbulence, large-scale structure 
formation in the universe, high-energetic multiparticle dynamics and
diffusion limited aggregation are but a few examples showing scaling
over a wide range.

A convenient tool to characterize the spatial configurations of these
processes is the multifractal formalism \cite{FED88}--\cite{FRI95}
which analyses the one-point statistics of singularity strengths under
the assumption of multiple local scale invariance. 
Typically, scale-invariance is a 
consequence of some hierarchical organisation of the underlying process.
Since this process occurs in space, 
its characterisation in terms of multifractals can be expected also to
yield information about spatial correlation functions.

Multivariate correlations, on the other hand, have long
been known to contain (at least in principle) complete information of the 
underlying random process. Moments and especially cumulants are
capable, in their multivariate form, of providing this information
by means of well-tested statistical procedures. As in the case of 
multifractals, cumulants can detect and characterize scale 
invariance, but unlike the former do not depend on its presence.

The existence of these two sets of tools naturally leads to the
question how they are interrelated:
given a selfsimilar process, does the multifractal
characterisation completely determine the spatial correlation functions
and vice versa, or does one provide more information than the other?

Some attempts have been made to elucidate the relationship between 
multifractals and correlation functions: two-point statistics of 
multifractal measures have been discussed in Refs.\ \cite{CAT87,MEN90}
and applied to turbulence in \cite{NEI93}. It is
to be expected, though, that only the full $n$-point statistics 
of singularity strengths would provide the equivalent information 
contained in the spatial correlation functions of all orders.

It is known that different selfsimilar processes can lead to
identical multifractal exponents \cite{FEI87,MEN87,SCH85}. 
This can be understood as a hint that the multifractal characterisation 
is indeed incomplete and that the spatial correlation functions 
may contain more information about the underlying process. 
We shall show that this is indeed the case for several examples of
multiplicative branching processes.

The ultimate goal for any analysis, of course, would be a
characterisation of the multivariate statistics gained 
from a physical process to  a degree of completeness 
that would permit precise reconstruction of
the ``branching rule'' that governs each step of the 
selfsimilar process. As far as theory is concerned, this 
holy grail would be achieved with the specification of a 
multivariate generating function whose $n$-fold derivatives 
would precisely reproduce the corresponding experimentally 
measured $n$-variate cumulants.

Clearly, finding this real-life generating function remains beyond
present-day capabilities, for both theoretical and experimental
reasons. It is possible, though, to make progress towards the ideal 
by inventing simple branching models whose generating functions can be 
calculated. An approach to calculate the $n$-point  spatial correlations 
to arbitrary order within some of these models was presented in 
Refs.\ \cite{GRE95,GRE96}: there, the multivariate generating function 
of the spatial correlations was constructed iteratively
from a backward evolution equation, leading to a recursive
derivation of spatial correlations. 
An important further step has been the recent discovery
of a large class of analytic solutions for generating functions
of selfsimilar multiplicative branching processes
\cite{GRE97b}.

In this paper, we expand on the latter discovery and attempt
to cast further light on said relationship between multifractals
and correlations.
After a brief explanation of the mechanics of multiplicative
branching processes in Sec.~\ref{sec-multip}, we review 
in Sec.~\ref{sptcor} the formalism of multivariate correlations and derive 
the abovementioned analytic solutions for the branching
generating functions.
While singularity strength variables are already treated in
Sec.~\ref{sptcor}, the relationship between spatial correlations
and multifractals is explored more fully in Sec.~\ref{mfrct}, 
where we also show how the vaunted splitting function 
(``branching rule'') can be
reconstructed via a two-dimensional Laplace transform.
The question of reconstructing this splitting function from data
is also briefly discussed.
Four examples of branching processes, treated in Sec.~\ref{xmpl},
drive home the message that multifractals contain less information
than the spatial correlations. We discuss our results
in more general terms in Sec.~\ref{concll}.

\section{Multiplicative branching processes}     
\label{sec-multip}

Random multiplicative branching processes, sometimes also called 
weight-curdling models or simply cascade models, can be
constructed for any number of branches per splitting. For simplicity,
we concentrate on binary processes; generalizations are straightforward.

The branching rule governing a binary cascade is described as follows:
a starting energy density $\epsilon^{(0)} = 1$, uniformly
distributed on the unit interval, is split up into two daughter
densities $\epsilon_0^{(1)} = q_0 \epsilon^{(0)}$ and 
$\epsilon_1^{(1)} = q_1 \epsilon^{(0)}$ living
on adjacent subintervals of length $2^{-1}$. The random weights 
$q_0$ and $q_1$, often also called multipliers or splitting parameters,
are drawn from a joint probability density (or ``splitting function'')
$p(q_0, q_1)$. In the next generation, each of the two daughter energy
densities is itself split up by the same branching rule into two 
granddaughters distributed uniformly over adjacent subintervals of length 
$2^{-2}$. Generally, energy densities of the $j$-th generation, 
$\epsilon^{(j)}$, are characterized by the binary index 
$\bkappa = (k_1 k_2 \cdots k_j)$,
with each $k$ taking on possible values 0 or 1. 
Successive repetition of this prescription for each $\epsilon$ of the
$j$-th generation yields, at the next branching, $2^{j+1}$ energy densities
$\epsilon_{k_1 \cdots k_j0}^{(j+1)} = 
 q_{k_1 \cdots k_j0}^{(j+1)} \epsilon_{k_1 \cdots k_j}^{(j)}$
and
$\epsilon_{k_1 \cdots k_j1}^{(j+1)} = 
 q_{k_1 \cdots k_j1}^{(j+1)} \epsilon_{k_1 \cdots k_j}^{(j)}$,
with the multipliers drawn from a splitting function
$p(q_{k_1 \cdots k_j0}^{(j+1)},q_{k_1 \cdots k_j1}^{(j+1)})$,
which usually (although not necessarily) is identical with 
$p(q_0, q_1)$. The process is completed after $J$ cascade steps.

When the splitting function $p(q_0,q_1)$
does not depend on the generation $j$ or branch location 
$k_1 \cdots k_j$, the above prescription gives rise to a selfsimilar 
multiplicative branching process, wholly characterized by the choice
of splitting function $p$. We list four examples for later use. 
The choice 
\begin{equation}
\label{21eins}
  p(q_0,q_1)
    =  {1 \over 2}
       \left[
       \rule{0mm}{1em}
       \delta ( q_0 - (1+\beta) ) + \delta ( q_0 - (1-\beta) )
       \right] \;
       \delta ( q_0 + q_1 -2)
\end{equation}
leads to the binomial multiplicative process, also known as the
$p$-model (often the equivalent parameter $p = ( 1+\beta )/2$ is used)
\cite{MEN87}. The delta function $\delta ( q_0 + q_1 -2)$ ensures
that this process conserves energy at every branching.

The splitting function for the non-energy conserving counterpart 
of the $p$-model, the $\alpha$-model \cite{SCH85}, is given by
\begin{equation}
\label{21zwei}
  p(q_0,q_1)
    =  \frac{1}{4} \prod_{k=0}^1
       \left[
       \rule{0mm}{1em}
       \delta ( q_k - (1+\beta) ) + \delta ( q_k - (1-\beta) )
       \right]
       \quad .
\end{equation}
For the energy dissipation process in fully developed turbulence, the
parametrisation
\begin{eqnarray}
\label{21drei}
  p(q_0,q_1)
    & = &  \frac{1}{2}
           \frac{\Gamma(2\beta)}{\Gamma(\beta)^2}
           \left( \frac{q_0}{2} \right)^{\beta-1}
           \left( \frac{q_1}{2} \right)^{\beta-1}
           \delta ( q_0 + q_1 -2)
           \nonumber\\
    & = &  \frac{1}{2^{2\beta-1}}
           \frac{\Gamma(2\beta)}{\Gamma(\beta)^2}
           \left( q_0 (2-q_0) \right)^{\beta-1}
           \delta ( q_0 + q_1 -2)
\end{eqnarray}
with $\beta = 3.2$ has been directly deduced from experiments
\cite{SRE95}. The corresponding splitting function not conserving energy,
is given by
\begin{equation}
\label{21vier}
  p(q_0,q_1)
    =  \left(
       \frac{1}{2^{2\beta-1}}
       \frac{\Gamma(2\beta)}{\Gamma(\beta)^2}
       \right)^2
       \left( q_0 (2-q_0) \right)^{\beta-1}
       \left( q_1 (2-q_1) \right)^{\beta-1}
       \quad .
\end{equation}
These four splitting functions are shown in Fig.\ 1.

\begin{figure}

\centerline{\epsfysize=150mm \epsfbox{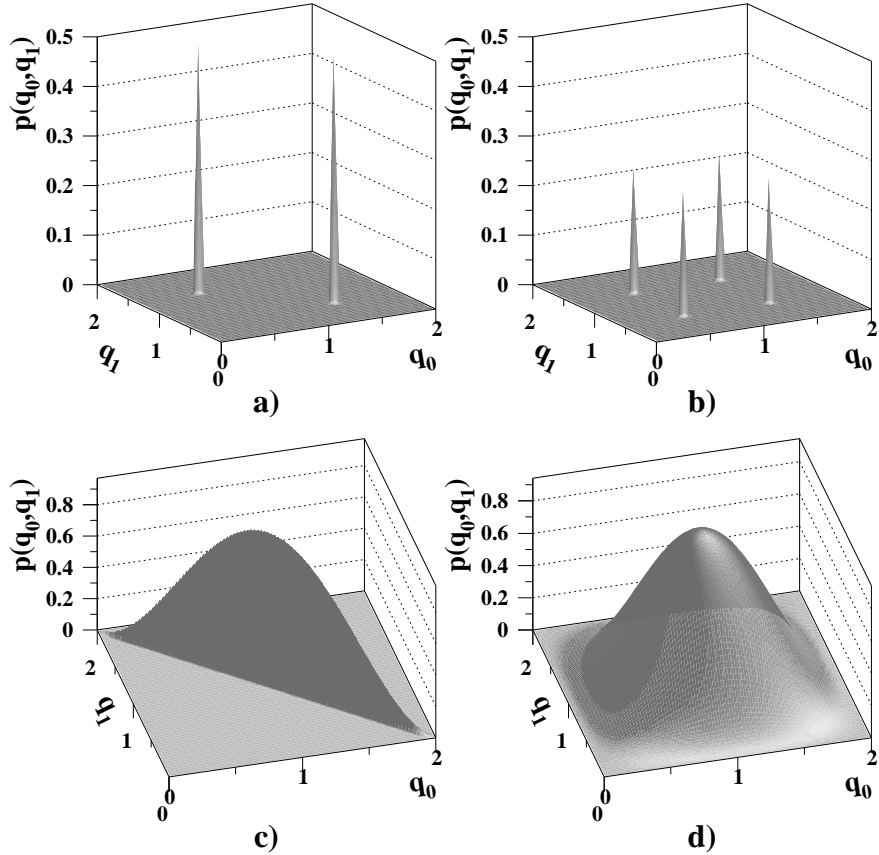}}

\caption{
Splitting functions $p(q_0,q_1)$ of
(a) the $p$-model with $\beta=0.4$,
(b) the $\alpha$-model with $\beta=0.4$,
(c) the energy-conserving SRST cascade model with $\beta=3.2$, and
(d) the energy-non-conserving SRST cascade model with $\beta=3.2$.
}
\label{fig1}
\end{figure}

\section{Spatial correlations}       
\label{sptcor}

\subsection{Possible variables}

For branching processes, the success in extracting useful information 
or presenting a clear picture depends strongly on the choice of
variable. In Refs.\ \cite{GRE95,GRE96,GIE97},
forward and backward evolution equations for a
multivariate generating function were employed to calculate the
correlations directly in terms of the energy densities,
$\langle \epsilon_{\bkappa_1}^{(J)} \cdots \epsilon_{\bkappa_n}^{(J)}
 \rangle$.
In the context of multifractals \cite{FED88}, on the other hand,
another set of variables, the so-called singularity strengths 
$\alpha_{\bkappa}^{(J)}$, are in use. The definition of these strengths,
$(2^{-J}) \epsilon_{\bkappa}^{(J)}  = (2^{-J})^{\alpha_{\bkappa}^{(J)}}$, 
means that they are related to the energy density variables 
$\epsilon_{\bkappa}^{(J)}$ by
\begin{equation}
\label{21fuenf}
  \alpha_{\bkappa}^{(J)}
    =  1 - \frac{1}{J \ln 2} \ln \epsilon_{\bkappa}^{(J)}
       \quad ,
\end{equation}
so that correlations
$\langle \alpha_{\bkappa_1}^{(J)} \cdots \alpha_{\bkappa_n}^{(J)}
 \rangle$
between the singularity strengths $\alpha_{\bkappa}^{(J)}$ are 
intimately related to corresponding correlations
$\langle (\ln\epsilon_{\bkappa_1}^{(J)}) \cdots 
 (\ln\epsilon_{\bkappa_n}^{(J)}) \rangle$
between the logarithms of the energy densities 
$\epsilon_{\bkappa}^{(J)}$.

We have shown previously \cite{GRE97b}
that correlations between the $\ln\epsilon$ lead to a particularly
transparent structure for multiplicative branching processes. 
Starting with correlations of the logarithms of
energy densities, we show below how this transparent structure
arises before translating these results into the language of
multifractals.

\subsection{Correlations and generating functions}

We consider the spatial correlations
\begin{equation}
\label{22eins}
  \rho_{\bkappa_1, \ldots, \bkappa_n}
    =  \left\langle
       \ln\epsilon_{\bkappa_1}^{(J)}
       \cdots
       \ln\epsilon_{\bkappa_n}^{(J)}
       \right\rangle
\end{equation}
of order $n$ between the variables $\ln\epsilon_{\bkappa}^{(J)}$.
The bracket $\langle \ldots \rangle$
indicates the averaging over all possible configurations.
Moments $\rho_{\bkappa_1, \ldots, \bkappa_n}$ are conveniently
calculated from the multivariate generating function  
\begin{equation} 
\label{22zwei}
  Z [ \lambda^{(J)} ]
    =  \left\langle 
       \exp\Biggl(
       \sum_{k_1, \ldots, \atop k_{J}=0}^1 
       \lambda_{k_1 \cdots k_J}^{(J)} 
       \ln\epsilon_{k_1 \cdots k_J}^{(J)}
       \Biggr)
       \right\rangle 
\end{equation} 
by taking appropriate derivatives with respect to the conjugate variables 
$\lambda_{\bkappa}^{(J)}$:
\begin{equation} 
\label{22drei}
  \rho_{\bkappa_1, \ldots, \bkappa_n}
    =  \left. 
       \frac{ \partial^n  Z [ \lambda^{(J)} ] }
            { \partial \lambda_{\bkappa_1}^{(J)}
              \cdots
              \partial \lambda_{\bkappa_n}^{(J)} }
       \right|_{ \lambda = 0 } 
       \quad . 
\end{equation} 

The corresponding multivariate cumulant generating function
\begin{equation}
\label{22vier}
  K [ \lambda^{(J)} ]
    =  \ln Z [ \lambda^{(J)} ]
\end{equation}
yields the cumulant correlation densities in the same way via
\begin{equation}
\label{22fuenf}
  C_{\bkappa_1, \ldots, \bkappa_n} 
    =  \left.
       {\partial^n K[\lambda^{(J)}] \over
        \partial\lambda_{\bkappa_1}^{(J)} \cdots
        \partial\lambda_{\bkappa_n}^{(J)}
       } 
       \right|_{\lambda^{(J)} = 0}
    =  \left\langle
       \ln\epsilon_{\bkappa_1}^{(J)}
       \cdots
       \ln\epsilon_{\bkappa_n}^{(J)}
       \right\rangle_{c}
       \quad ;
\end{equation}
here, we have introduced the index $c$ (for ``cumulant'') on the last 
bracket to distinguish (\ref{22fuenf}) from the expression 
(\ref{22eins}) for the ordinary correlation densities. The lowest
four orders of cumulant correlation densities are
\begin{eqnarray}
\label{22sechs}
  C_{\bkappa_1} 
    &=&  \langle \ln\epsilon_{\bkappa_1} \rangle_c
         =  \langle \ln\epsilon_{\bkappa_1} \rangle 
         \quad ,
         \\
\label{22sieben}
  C_{\bkappa_1, \bkappa_2}
    &=&  \langle 
         \ln\epsilon_{\bkappa_1} \ln\epsilon_{\bkappa_2}
         \rangle_c
         \nonumber \\
    &=&  \langle  
         \ln\epsilon_{\bkappa_1} \ln\epsilon_{\bkappa_2}
         \rangle
         -   
         \langle \ln\epsilon_{\bkappa_1} \rangle
         \langle \ln\epsilon_{\bkappa_2} \rangle
         \quad ,
         \\
\label{22acht}
  C_{\bkappa_1, \bkappa_2, \bkappa_3}
     &=&  \langle  
          \ln\epsilon_{\bkappa_1} \ln\epsilon_{\bkappa_2}
          \ln\epsilon_{\bkappa_3}
          \rangle_c
          \nonumber \\
     &=&  \langle 
          \ln\epsilon_{\bkappa_1} \ln\epsilon_{\bkappa_2}
          \ln\epsilon_{\bkappa_3}
          \rangle
          - 
          \sum_{\{3\}}
          \langle 
          \ln\epsilon_{\bkappa_1} \ln\epsilon_{\bkappa_2}
          \rangle
          \langle \ln\epsilon_{\bkappa_3} \rangle
          \nonumber \\
     & &  + 2
          \langle \ln\epsilon_{\bkappa_1} \rangle
          \langle \ln\epsilon_{\bkappa_2} \rangle
          \langle \ln\epsilon_{\bkappa_3} \rangle ,
         \\
\label{22neun}
  C_{\bkappa_1, \bkappa_2, \bkappa_3, \bkappa_4}
     &=&  \langle  
          \ln\epsilon_{\bkappa_1} \ln\epsilon_{\bkappa_2}
          \ln\epsilon_{\bkappa_3} \ln\epsilon_{\bkappa_4}
          \rangle_c
          \nonumber \\
     &=&  \langle 
          \ln\epsilon_{\bkappa_1} \ln\epsilon_{\bkappa_2}
          \ln\epsilon_{\bkappa_3} \ln\epsilon_{\bkappa_4}
          \rangle
          - 
          \sum_{\{4\}}
          \langle 
          \ln\epsilon_{\bkappa_1} \ln\epsilon_{\bkappa_2}
          \ln\epsilon_{\bkappa_3}
          \rangle
          \langle \ln\epsilon_{\bkappa_4} \rangle
          \nonumber \\
     & &  -
          \sum_{\{3\}}
          \langle 
          \ln\epsilon_{\bkappa_1} \ln\epsilon_{\bkappa_2}
          \rangle
          \langle 
          \ln\epsilon_{\bkappa_3} \ln\epsilon_{\bkappa_4} 
          \rangle
          + 
          2 \sum_{\{6\}}
          \langle 
          \ln\epsilon_{\bkappa_1} \ln\epsilon_{\bkappa_2}
          \rangle
          \langle \ln\epsilon_{\bkappa_3} \rangle
          \langle \ln\epsilon_{\bkappa_4} \rangle
          \nonumber \\
     & &  - 6
          \langle \ln\epsilon_{\bkappa_1} \rangle
          \langle \ln\epsilon_{\bkappa_2} \rangle
          \langle \ln\epsilon_{\bkappa_3} \rangle
          \langle \ln\epsilon_{\bkappa_4} \rangle
          \quad ,
\end{eqnarray}
with brackets $\{ \ldots \}$ indicating summation over permutations
of indices.

\subsection{Construction of the cumulant generating function
               (first approach)}

The derivation of the cumulant generating function depends crucially
on a rearrangement of terms in the sum
$\sum_{k_1, \ldots, k_J} \lambda_{k_1 \cdots k_J}^{(J)}
                    \ln\epsilon_{k_1 \cdots k_J}^{(J)}
$ 
entering the generating function in (\ref{22zwei}). The trick lies in
recognising that
the cascade prescription of Section \ref{sec-multip} implies
that the energy density in the $J$-th generation is the product
of all the splitting parameters in its ancestry,
\begin{equation}
\label{221zwei}
  \epsilon_{\bkappa}^{(J)}
    =  \epsilon_{k_1 \cdots k_J}^{(J)} 
    =  q_{k_1}^{(1)} 
       q_{k_1 k_2}^{(2)} \cdots
       q_{k_1 \cdots k_J}^{(J)}
       \quad ,
\end{equation}
so that its logarithm is additive in the multipliers' logarithms,
\begin{equation}
\label{221drei}
  \ln\epsilon_{k_1 \cdots k_J}^{(J)} 
    =  \sum_{j=1}^J \ln q_{k_1 \cdots k_j}^{(j)}
       \quad .
\end{equation}
Inserting this into the sum in the exponential of the generating
function in (\ref{22zwei}) and defining 
\begin{equation}
\label{221sieben}
\lambda_{k_1\cdots k_j}^{(j)} 
\equiv 
\sum_{k_{j+1}, \ldots, k_J} 
     \lambda_{k_1\cdots k_j k_{j+1} \cdots k_J}^{(J)}
   \,,
\end{equation}
we find, after rearrangement and judicious regrouping of terms, that
\begin{eqnarray}
\label{hceb}
\sum_{k_1, \ldots, k_J} \lambda_{k_1 \cdots k_J}^{(J)}
                    \ln\epsilon_{k_1 \cdots k_J}^{(J)}
&=&
\sum_{j=1}^J 
\sum_{k_1, \ldots, k_j}
          \lambda_{k_1 \cdots k_j}^{(j)}
            \ln q_{k_1 \cdots k_j}^{(j)}
\\
&=&
\sum_{j=1}^J \sum_{{k_1,\ldots, \atop k_{j-1}}}
\left(
  \lambda_{k_1 \cdots k_{j-1}0}^{(j)} \ln q_{k_1 \cdots k_{j-1}0}^{(j)}
+ \lambda_{k_1 \cdots k_{j-1}1}^{(j)} \ln q_{k_1 \cdots k_{j-1}1}^{(j)}
\right)
\,.
\nonumber
\end{eqnarray}

This can now be utilized to rewrite the 
configuration average $\langle {\cal O}(\epsilon^{(J)}) \rangle$
of any function $\cal O$ in terms of the splitting parameters as
follows. We start with the definition
\begin{equation}
\label{221eins}
  \langle {\cal O}(\epsilon^{(J)}) \rangle
    =  \int 
       d(\ln\epsilon_{0\cdots 0}^{(J)})
       \cdots
       d(\ln\epsilon_{1\cdots 1}^{(J)}) \;
       p( \ln\epsilon_{0\cdots 0}^{(J)},
          \cdots ,
          \ln\epsilon_{1\cdots 1}^{(J)} ) \;
       {\cal O}(\epsilon^{(J)})
       \quad ,
\end{equation}
where 
$p( \ln\epsilon_{0\cdots 0}^{(J)}, \cdots , 
 \ln\epsilon_{1\cdots 1}^{(J)} )$ 
is the joint probability of finding a given logarithmic energy density
$\ln\epsilon_{0\cdots 0}^{(J)}$ in bin ($0\cdots 0$), a given
$\ln\epsilon_{0\cdots 1}^{(J)}$ in the next bin, etc.
Since each energy density $\epsilon_{\bkappa}^{(J)}$
is completely determined by its $J$ ancestral splitting
parameters, the multivariate probability density
$p( \ln\epsilon_{0\cdots 0}^{(J)}, \cdots ,  
 \ln\epsilon_{1\cdots 1}^{(J)} )$
can be constructed from the splitting function $p(q_0,q_1)$
by using (\ref{221drei}):
\begin{eqnarray}
\label{221vier}
  p(\ln\epsilon_{0\cdots 0}^{(J)}, \ldots, \ln\epsilon_{1\cdots 1}^{(J)})
    &=&  \int \Bigl[ \prod_{j=1}^J 
         \prod_{  {k_1, \ldots, \atop k_{j-1}=0 }  }^1
         dq_{k_1 \cdots k_{j-1} 0}^{(j)} \,
         dq_{k_1 \cdots k_{j-1} 1}^{(j)} \,
         p ( q_{k_1 \cdots k_{j-1} 0}^{(j)},
             q_{k_1 \cdots k_{j-1} 1}^{(j)}  )
         \Bigr] 
         \nonumber \\
    & &  \qquad \times
         \Biggl[
         \prod_{  {k_1, \ldots, \atop k_{J}=0 }  }^1
         \delta \Bigl( 
         \ln\epsilon_{k_1 \cdots k_J}^{(J)}
         -  
         \sum_{j=1}^J \ln q_{k_1 \cdots k_j}^{(j)}
         \Bigr)
         \Biggr]
         \quad .
\end{eqnarray}
Inserting (\ref{hceb}), (\ref{221eins}) and (\ref{221vier})
into Eq.\ (\ref{22vier}), we find that the cumulant generating function 
becomes
\begin{eqnarray}
\label{221fuenf}
  K [ \lambda_{0\cdots 0}^{(J)} , \ldots, \lambda_{1\cdots 1}^{(J)} ]
    &=&  \ln\Biggl[ \int \Bigl(
         \prod_{  {k_1, \ldots, \atop k_{J}=0 }  }^1
         d(\ln \epsilon_{k_1\cdots k_J}^{(J)} )
         \Bigr) \;
         p(\ln\epsilon_{0\cdots 0}^{(J)}, \ldots, 
           \ln\epsilon_{1\cdots 1}^{(J)})
         \Biggr.
         \nonumber \\
    & &  \Biggl. \qquad \times
         \exp \Biggl( 
         \sum_{k_1,\ldots, \atop k_J = 0}^1 
         \lambda_{k_1\cdots k_J}^{(J)} 
         \ln \epsilon_{k_1\cdots k_J}^{(J)} 
         \Biggr)
         \Biggr]
         \nonumber \\
    &=&  \sum_{j=1}^J 
         \sum_{k_1,\ldots, \atop k_{j-1} = 0}^1
         Q[ \lambda_{k_1\cdots k_{j-1} 0}^{(j)},
            \lambda_{k_1\cdots k_{j-1} 1}^{(j)} ]
         \quad,
\end{eqnarray}
where
\begin{eqnarray}
\label{221sechs}
  Q[ \lambda_{0}, \lambda_{1} ]
    &=&  \ln \Bigl[ 
         \int dq_{0}\, dq_{1}\, p(q_{0}, q_{1})
         \exp \left(
         \lambda_{0} \ln q_{0} 
         + 
         \lambda_{1} \ln q_{1}
         \right) 
         \Bigr]
         \nonumber \\  
    &=&  \ln \Bigl\langle
         \exp \left(
         \lambda_{0} \ln q_{0} 
         + 
         \lambda_{1} \ln q_{1}
         \right) 
         \Bigr\rangle 
\end{eqnarray}
Eqs.\ (\ref{221fuenf})--(\ref{221sechs})
represent the long-sought analytic expression
for multiplicative cascades: they show that
the cumulant generating function of the entire cascade can be 
written as the sum of all branching generating functions $Q$, one
for every branching. In most cases, the two-fold integrals
(\ref{221sechs}) can be solved analytically or parametrically
to yield a complete analytic solution for $K$ and thereby for
all cumulants.

The scope and limitations of the solution (\ref{221fuenf}) are as
follows. Clearly, the generating function in terms of
$\ln\epsilon$ is applicable to any functional form of the
splitting function or b.g.f. It does not depend on the number of
branches either: trivariate or even higher-variate splitting
functions can be implemented. Due to the additive nature of
the b.g.f.'s, the splitting functions can differ from
generation to generation and even from branch to branch.
The only (and important) precondition for the applicability
of Eq.\ (\ref{221fuenf}) is that the splitting variables
of every branching must be independent of those of the other branchings
and generations.

\subsection{Construction of the cumulant generating function
               (second approach)}

The cumulant generating function can also be constructed from a 
forward evolution equation.
In Ref.\ \cite{GIE97}, the forward evolution equation 
for the generating function was derived in terms of energy densities 
$\epsilon_{k_1 \cdots k_j}^{(j)}$. In terms of the
$\ln\epsilon_{k_1 \cdots k_j}^{(j)}$,
the cumulant generating function after $j$ cascade steps,
$K^{(j)} [ \lambda^{(j)} ]$, can be expressed in terms of
$K^{(j-1)} [ \lambda^{(j-1)} ]$ as follows.
At each of the $2^{j-1}$ independent branchings, Eq.\  (\ref{221drei})
relates the daughter energy densities to their parent by
\begin{equation}
\label{222eins}
  \ln\epsilon_{k_1 \cdots k_j}^{(j)} 
    =  \ln\epsilon_{k_1 \cdots k_{j-1}}^{(j-1)} 
       + 
       \ln q_{k_1 \cdots k_j}^{(j)} 
       \qquad ( k_j = 0,1) \,.
\end{equation}
In analogy to (\ref{hceb}),
\begin{eqnarray} 
\label{222zwei}
  \lefteqn{ 
  \exp\Biggl(  
  \sum_{k_1, \ldots , \atop k_j = 0}^1 
  \lambda_{k_1 \cdots k_j}^{(j)} 
  \ln\epsilon_{k_1 \cdots k_j}^{(j)}
  \Biggr) = 
  }  \nonumber  \\ 
    &=&  \prod_{k_1, \ldots , \atop k_{j-1} = 0}^1
         \exp\left(   
         \lambda_{k_1 \cdots k_{j-1} 0}^{(j)} 
         \ln\epsilon_{k_1 \cdots k_{j-1} 0}^{(j)}
         + 
         \lambda_{k_1 \cdots k_{j-1} 1}^{(j)} 
         \ln\epsilon_{k_1 \cdots k_{j-1} 1}^{(j)}
         \right) 
         \nonumber \\ 
    &=&  \prod_{k_1, \ldots , \atop k_{j-1} = 0}^1
         \exp\left[ 
         \left(
         \lambda_{k_1 \cdots k_{j-1} 0}^{(j)}
         +
         \lambda_{k_1 \cdots k_{j-1} 1}^{(j)} 
         \right)
         \ln\epsilon_{k_1 \cdots k_{j-1}}^{(j-1)}
         \right.
         \nonumber \\
    & &  \qquad\qquad \left. +
         \lambda_{k_1 \cdots k_{j-1} 0}^{(j)} 
         \ln q_{k_1 \cdots k_{j-1} 0}^{(j)} 
         +
         \lambda_{k_1 \cdots k_{j-1} 1}^{(j)} 
         \ln q_{k_1 \cdots k_{j-1} 1}^{(j)} 
         \right]
\end{eqnarray}
and assigning, as in (\ref{221sieben}),
\begin{equation} 
\label{222drei}
  \lambda_{k_1 \cdots k_{j-1} 0}^{(j)} 
  +
  \lambda_{k_1 \cdots k_{j-1} 1}^{(j)}
    =  \lambda_{k_1 \cdots k_{j-1}}^{(j-1)}
       \quad ,
\end{equation}
we arrive at 
\begin{eqnarray} 
\label{222vier}
  K^{(j)} [ \lambda^{(j)} ]
    &=&  K^{(j-1)} [ \lambda^{(j-1)} ]
         \nonumber \\
    & &  +
         \sum_{k_1,\ldots, \atop k_{j-1} = 0}^1
         \ln \Bigl[ \int 
         dq_{k_1\cdots k_{j-1} 0}^{(j)}\, 
         dq_{k_1\cdots k_{j-1} 1}^{(j)}\;  
         p(q_{k_1\cdots k_{j-1} 0}^{(j)}, q_{k_1\cdots k_{j-1} 1}^{(j)}) 
         \Bigr.
         \nonumber \\
    & &  \qquad\qquad\qquad \Bigl.
         \times \exp 
         \left(
         \lambda_{k_1\cdots k_{j-1} 0}^{(j)} 
         \ln q_{k_1\cdots k_{j-1} 0}^{(j)} 
         + 
         \lambda_{k_1\cdots k_{j-1} 1}^{(j)} 
         \ln q_{k_1\cdots k_{j-1} 1}^{(j)}
         \right)
         \Bigr] 
         \nonumber \\
    &=&  K^{(j-1)} [ \lambda^{(j-1)} ]
         +
         \sum_{k_1,\ldots, \atop k_{j-1} = 0}^1
         Q[ \lambda_{k_1\cdots k_{j-1} 0}^{(j)},
            \lambda_{k_1\cdots k_{j-1} 1}^{(j)} ]
         \quad .
\end{eqnarray}
This forward evolution equation can be iterated from $j=J$ down to
$j=0$, where we arrive at
$K^{(j=0)} [ \lambda^{(0)} ]
   =  \ln ( Z^{(0)} [ \lambda^{(0)} ] )
   =  0 $;
the outcome of this iteration is identical to the solution 
(\ref{221fuenf}).
Similarly, a backward evolution equation for the cumulant generating functions
can also be used to achieve identical results; we leave this as
an exercise to the reader.

\subsection{Cumulant correlation densities}
\label{cumcord}

Having found an analytic expression of the multivariate generating 
function $K [ \lambda^{(J)} ]$, it is now straightforward to
calculate the cumulant correlation densities 
(\ref{22sechs})--(\ref{22neun}) between the
$\ln\epsilon_{\bkappa}^{(J)}$-variables via Eq.\ (\ref{22fuenf}). 
In order to give compact expressions, we define
an ultrametric distance between bins.
Assume the two bins are together for the first $j$ steps of the binary 
cascade before splitting at the $(j+1)$-th generation, 
$\bkappa_1 = ( k_1 \cdots k_j k_{j+1} \cdots k_J )$
and 
$\bkappa_2 = ( k_1 \cdots k_j k_{j+1}^\prime \cdots k_J^\prime )$,
where $k_i = k_i^\prime$ for all $1 \leq i \leq j$ and 
$k_{j+1} \neq k_{j+1}^\prime$. 
Then the ultrametric distance between them is
\begin{equation} 
\label{223eins}
  d_2
    =  {\rm dist}(\bkappa_1,\bkappa_2)
    =  J - j
       \quad .
\end{equation}
For $n$ bins, the generalised ultrametric distance is
\begin{equation} 
\label{223zwei}
  d_n
    =  \max_{1 \leq i < j \leq n}
       {\rm dist}(\bkappa_i,\bkappa_j)
       \quad .
\end{equation}

In the following, we assume that the same functional form 
for the b.g.f.\ is used at all branchings.
We also assume the splitting function $p(q_0,q_1)$ to be
symmetric under exchange of the splitting parameters,
$p(q_0,q_1) = p(q_1,q_0)$; the four splitting functions
given in Eqs.\ (\ref{21eins})--(\ref{21vier}) fall into this category. 
In consequence, the branching generating function
$Q [ \lambda_0 , \lambda_1 ] = Q [ \lambda_1 , \lambda_0 ]$
is also symmetric, so that also
\begin{equation} 
\label{223drei}
  \left.
  \frac{ \partial^{n_1+n_2} Q[\lambda_0,\lambda_1] }
       { \partial\lambda_0^{n_1} \partial\lambda_1^{n_2} }
  \right|_{\lambda=0}
    =  \left.
       \frac{ \partial^{n_1+n_2} Q[\lambda_0,\lambda_1] }
            { \partial\lambda_0^{n_2} \partial\lambda_1^{n_1} }
       \right|_{\lambda=0}
       \quad .
\end{equation}
This relation results in compact expressions for the 
cumulant correlation densities $C_{\bkappa_1, \ldots, \bkappa_n}$.
Inserting (\ref{221fuenf}) into Eq.\ (\ref{22fuenf})
and taking into account (\ref{221sieben}),
the cumulant correlation density of first order is found to be
\begin{equation} 
\label{223vier}
  C_{\bkappa_1}
    =  J \left( \left.
       \frac{ \partial Q[\lambda_0,\lambda_1] }
            { \partial\lambda_0 }
       \right|_{\lambda=0}
       \right)
       \quad .
\end{equation}
For second order we get, with (\ref{223eins}),
\begin{equation} 
\label{223fuenf}
  C_{\bkappa_1,\bkappa_2}
    =  (J-d_2) \left( \left.
       \frac{ \partial^2 Q[\lambda_0,\lambda_1] }
            { \partial\lambda_0^2 }
       \right|_{\lambda=0}
       \right)
       +
       ( 1 - \delta_{d_2,0} )
       \left( \left.
       \frac{ \partial^2 Q[\lambda_0,\lambda_1] }
            { \partial\lambda_0 \partial\lambda_1 }
       \right|_{\lambda=0}
       \right)
       \quad ,
\end{equation}
while with the generalisation (\ref{223zwei}), the third order cumulant
density becomes
\begin{equation} 
\label{223sechs}
  C_{\bkappa_1,\bkappa_2,\bkappa_3}
    =  (J-d_3) \left( \left.
       \frac{ \partial^3 Q[\lambda_0,\lambda_1] }
            { \partial\lambda_0^3 }
       \right|_{\lambda=0}
       \right)
       +
       ( 1 - \delta_{d_3,0} )
       \left( \left.
       \frac{ \partial^3 Q[\lambda_0,\lambda_1] }
            { \partial\lambda_0^2 \partial\lambda_1 }
       \right|_{\lambda=0}
       \right)
       \quad .
\end{equation}
In the case of fourth order, two cases have to be
distinguished, depending on whether the first splitting of the four
bins $\bkappa_1, \ldots , \bkappa_4$ goes into three and one, say 
($\bkappa_1\bkappa_2\bkappa_3 | \bkappa_4$)
with $d_3(\bkappa_1,\bkappa_2,\bkappa_3) < d_4$ and $n_1=3, n_2=1$,
or into two and two, say
($\bkappa_1\bkappa_2 | \bkappa_3\bkappa_4$)
with $d_2(\bkappa_1,\bkappa_2) < d_4$, $d_2(\bkappa_3,\bkappa_4) < d_4$
and $n_1=2, n_2=2$:
\begin{equation} 
\label{223sieben}
  C_{\bkappa_1,\bkappa_2,\bkappa_3,\bkappa_4}
    =  (J-d_4) \left( \left.
       \frac{ \partial^4 Q[\lambda_0,\lambda_1] }
            { \partial\lambda_0^4 }
       \right|_{\lambda=0}
       \right)
       +
       ( 1 - \delta_{d_4,0} )
       \left( \left.
       \frac{ \partial^4 Q[\lambda_0,\lambda_1] }
            { \partial\lambda_0^{n_1} \partial\lambda_1^{n_2} }
       \right|_{\lambda=0}
       \right)
       \quad .
\end{equation}
Even under the stated assumptions, 
the expressions (\ref{223vier})--(\ref{223sieben}) are still very general
for selfsimilar binary multiplicative cascade models. The one and only
input is the splitting function $p(q_0,q_1)$ determining the 
branching generating function $Q[\lambda_0,\lambda_1]$.
The cumulant correlation densities for the splitting functions 
(\ref{21eins})--(\ref{21vier}) will be discussed in more detail
in Sec.\ \ref{xmpl}.

It is clear from Eqs.\ (\ref{223vier})--(\ref{223sieben}) that the
derivatives of the branching generating function completely fix the 
spatial cumulant correlation densities. Using the definition 
(\ref{221sechs}) these derivatives can be expressed in terms of
generic branching moments:
\begin{equation} 
\label{223acht}
  \left.
  \frac{ \partial^{n_1+n_2} Q[\lambda_0,\lambda_1] }
       { \partial\lambda_0^{n_1} \partial\lambda_1^{n_2} }
  \right|_{\lambda=0}
    =  \left\langle
       (\ln q_0)^{n_1} (\ln q_1)^{n_2}
       \right\rangle_c
       \quad .
\end{equation}
In the lowest three orders, they read explicitly:
\begin{eqnarray}
\label{223neun}
  \left\langle \ln q_0 \right\rangle_c
    &=&  \left\langle \ln q_0 \right\rangle
         \quad ,
         \nonumber \\
  \left\langle (\ln q_0)^2 \right\rangle_c
    &=&  \left\langle (\ln q_0)^2 \right\rangle
         -
         \left\langle \ln q_0 \right\rangle^2
         \quad ,
         \nonumber \\
  \left\langle (\ln q_0) (\ln q_1) \right\rangle_c
    &=&  \left\langle (\ln q_0) (\ln q_1) \right\rangle
         -
         \left\langle \ln q_0 \right\rangle
         \left\langle \ln q_1 \right\rangle
         \quad ,
         \nonumber \\
  \left\langle (\ln q_0)^3 \right\rangle_c
    &=&  \left\langle (\ln q_0)^3 \right\rangle
         - 3
         \left\langle (\ln q_0)^2 \right\rangle
         \left\langle \ln q_0 \right\rangle
         + 2
         \left\langle \ln q_0 \right\rangle^3
         \quad ,
         \nonumber \\
  \left\langle (\ln q_0)^2 (\ln q_1) \right\rangle_c
    &=&  \left\langle (\ln q_0)^2 (\ln q_1) \right\rangle
         - 2
         \left\langle (\ln q_0) (\ln q_1) \right\rangle
         \left\langle \ln q_0 \right\rangle
         -
         \left\langle (\ln q_0)^2 \right\rangle
         \left\langle \ln q_1 \right\rangle
         \nonumber \\
    & &  + 2
         \left\langle \ln q_0 \right\rangle^2
         \left\langle \ln q_1 \right\rangle
         \quad .
\end{eqnarray}

\subsection{Link to singularity strength variables}

We have defined the cumulant correlation densities
$C_{\bkappa_1, \ldots, \bkappa_n}
 = \langle \ln\epsilon_{\bkappa_1}^{(J)} \cdots
   \ln\epsilon_{\bkappa_n}^{(J)} \rangle_c$
in terms of the $\ln\epsilon_{\bkappa}^{(J)}$-variables. With the help of
(\ref{21fuenf}), they can be transformed into cumulant correlation densities
$C_{\bkappa_1, \ldots, \bkappa_n}^\alpha
 = \langle \alpha_{\bkappa_1}^{(J)} \cdots
   \alpha_{\bkappa_n}^{(J)} \rangle_c$
between the singularity strength variables
$\alpha_{\bkappa}^{(J)}$. We find for the corresponding generating function
\begin{eqnarray} 
\label{224eins}
  K^\alpha [\lambda^{(J)}]
    &=&  \ln \left\langle
         \exp\Biggl(
         \sum_{k_1, \ldots , \atop k_J = 0}^1
         \lambda_{k_1 \cdots k_J}^{(J)}
         \alpha_{k_1 \cdots k_J}^{(J)}
         \Biggr)
         \right\rangle
         \nonumber \\
    &=&  \ln \left\langle
         \exp\Biggl(
         \sum_{k_1, \ldots , \atop k_J = 0}^1
         \lambda_{k_1 \cdots k_J}^{(J)}
         \left( 
         1
         -
         \frac{1}{J \ln 2} \ln\epsilon_{k_1 \cdots k_J}^{(J)}
         \right)
         \Biggr)
         \right\rangle
         \nonumber \\
    &=&  \sum_{k_1, \ldots , \atop k_J = 0}^1
         \lambda_{k_1 \cdots k_J}^{(J)}
         +
         K \left[ \frac{(-1)}{J \ln 2} \lambda^{(J)} \right]
         \quad .
\end{eqnarray}
Taking derivatives, the relations between the cumulant densities
are found to be, for first order,
\begin{equation} 
\label{224zwei}
  C_{\bkappa_1}^\alpha
    =  \langle \alpha_{\bkappa_1}^{(J)} \rangle_c
    =  1 
       -
       \frac{1}{J \ln 2}
       \langle \ln\epsilon_{\bkappa_1}^{(J)} \rangle_c
    =  1 
       -
       \frac{1}{J \ln 2}
       C_{\bkappa_1}
       \quad ,
\end{equation}
and, for $n\geq 2$,
\begin{equation} 
\label{224drei}
  C_{\bkappa_1, \ldots, \bkappa_n}^\alpha
    =  \langle 
       \alpha_{\bkappa_1}^{(J)} \cdots \alpha_{\bkappa_n}^{(J)}
       \rangle_c
    =  \left( \frac{-1}{J \ln 2} \right)^n
       \langle 
       \ln\epsilon_{\bkappa_1}^{(J)} \cdots \ln\epsilon_{\bkappa_n}^{(J)}
       \rangle_c
    =  \left( \frac{-1}{J \ln 2} \right)^n
       C_{\bkappa_1, \ldots, \bkappa_n}
       \,,
\end{equation}
in other words, with the exception of first order,
the $C_{\bkappa_1, \ldots, \bkappa_n}^\alpha$
are directly proportional to $C_{\bkappa_1, \ldots, \bkappa_n}$.

\section{Spatial correlations vs.\ multifractals} 
\label{mfrct}

\subsection{Multifractals}

One way to approach multifractals is to count how often specific
values $\alpha$ of the singularity strengths occur; this leads to
the $f(\alpha)$-spectrum
\cite{FED88}.
The latter is related by a Legendre transformation to a sequence of mass
exponents $\tau (\nu)$ which are defined by the moment scaling behavior
of bin energies $E_{k_1 \cdots k_j}^{(j)}$:
\begin{equation}
\label{31eins}
  M_\nu(j)
    =  \left\langle
       \sum_{k_1, \ldots , \atop k_j = 0}^1  
       \left( E_{k_1 \cdots k_j}^{(j)} \right)^\nu
       \right\rangle
    =  \left\langle
       \sum_{k_1, \ldots , \atop k_j = 0}^1  
       \left( 
       \frac{\epsilon_{k_1 \cdots k_j}^{(j)}}{2^j} 
       \right)^\nu
       \right\rangle
    =  (2^{-j})^{- \tau (\nu)}
       \quad .
\end{equation}
In order to avoid confusion we have written $\tau(\nu)$
instead of the more familiar notation $\tau(q)$.
Since the bin energies are related to the singularity strengths by
$E_{\bkappa}^{(j)} = (2^{-j})^{\alpha_{\bkappa}^{(j)}}$,
these moments are to be understood as scale-dependent measures for 
the one-point statistics of the singularity strengths.

We first derive the expression for the exponents $\tau (\nu)$ for 
the special case of a symmetric splitting function 
$p(q_0,q_1) = p(q_1,q_0)$ of a binary multiplicative
cascade process which conserves energy. (Note that 
for energy non-conserving splitting functions, as for
example those given in (\ref{21zwei}) and (\ref{21vier}),
the multifractal exponents cannot be defined unambiguously; see
Ref.\ \cite{GRE96}.)
Then we can use (\ref{221zwei}) and the statistical independence
of splitting parameters to deduce
\begin{equation}
\label{31zwei}
  \left\langle 
  \left(   \epsilon_{k_1 \cdots k_j}^{(j)}   \right)^\nu 
  \right\rangle
    =  \langle q_{k_1}^\nu \rangle
       \langle q_{k_1 k_2}^\nu \rangle
       \cdots
       \langle q_{k_1 \cdots k_j}^\nu \rangle
    =  \langle  q_0^\nu  \rangle^j
       \quad .
\end{equation}
Insertion into Eq.\ (\ref{31eins}) yields:
\begin{equation}
\label{31drei}
  \tau (\nu)
    =  \frac{1}{\ln 2}
       \ln\left(
       \frac{ \left\langle  q_0^\nu  \right\rangle }{ 2^{\nu -1} }
       \right)
    =  -(\nu-1)
       + \frac { \ln \left\langle  q_0^\nu  \right\rangle }
               { \ln 2 }
       \quad .
\end{equation}
For a $q_0$/$q_1$-asymmetric, energy-conserving splitting function 
we have to replace the sum appearing in (\ref{31eins}) by
\begin{equation}
\label{31vier}
  \sum_{k_1, \ldots , \atop k_j = 0}^1
  \left\langle \left( 
  E_{k_1 \cdots k_j}^{(j)} 
  \right)^\nu \right\rangle
    =  \frac{1}{2^{\nu j}}
       \left[
       \left\langle  q_0^\nu  \right\rangle
       + \left\langle  q_1^\nu  \right\rangle
       \right]^j
       \quad ,
\end{equation}
which gives rise to
\begin{equation}
\label{31fuenf}
  \tau (\nu)
    =  \frac{1}{\ln 2}
       \ln\left(
       \frac{ \left\langle  q_0^\nu  \right\rangle
              + \left\langle  q_1^\nu  \right\rangle }{ 2^\nu }
       \right)
    =  -\nu
       + \frac{ \ln\left(
                \left\langle  q_0^\nu  \right\rangle
                + \left\langle  q_1^\nu  \right\rangle
                    \right) }
              { \ln 2 }
           \quad .
\end{equation}

\subsection{Relationship between $Q[\lambda_0,\lambda_1]$ and 
            $\tau (\nu)$}

The close relationship between the branching generating function
$Q[\lambda_0,\lambda_1]$ and the multifractal exponents
$\tau (\nu)$ is revealed by simple inspection of
(\ref{221sechs}) and (\ref{31fuenf}). We find that
\begin{equation}
\label{32eins}
  \tau (\nu)
    =  -\nu
       + \frac{1}{\ln 2}
       \ln\Bigl[
       \exp\left( Q[\lambda_0=\nu,\lambda_1=0] \right)
       + \exp\left( Q[\lambda_0=0,\lambda_1=\nu] \right)
       \Bigr]
       \quad .
\end{equation}
This relationship\footnote{
The univariate version of this relation has already been 
discussed by Novikov \cite{NOV71} in connection with the statistics
of generalised multipliers, the so-called breakdown coefficients.
}
simplifies further once we consider an 
energy-conserving, $q_0 / q_1$ symmetric splitting function.
Through (\ref{31drei}) we arrive at
\begin{equation}
\label{32zwei}
  \tau (\nu)
    =  -(\nu-1)
       + \frac{1}{\ln 2} Q[\lambda_0=\nu,\lambda_1=0]
       \quad .
\end{equation}

The two relations (\ref{32eins}) and (\ref{32zwei})
express $\tau (\nu)$ in terms of $Q[\lambda_0,\lambda_1]$.
It appears that $\tau$ is more limited than $Q$ in that the latter
is defined over the full $\lambda_0$--$\lambda_1$ plane while
$\tau(\nu)$ is restricted to the axes $\lambda_0=0$ or $\lambda_1=0$.

Let us consider the point in more detail.
In particular, we will demonstrate that different
energy-conserving splitting functions may lead to identical multifractal
exponents while their branching generating functions do differ.

Given an energy-conserving, $q_0/q_1$-symmetric splitting function 
$p(q_0,q_1) = p(q_1,q_0) = p(q_0) \delta (q_0+q_1-2)$,
we construct a related splitting function by breaking the 
$q_0/q_1$-symmetry:
\begin{equation}
\label{32vier}
  \tilde{p}(q_0,q_1)
    =  \left\{
       \begin{array}{ll}
       2p(q_0,q_1)
         &  \qquad (0 \leq q_0 \leq 1 \leq q_1 \leq2) \\
       0
         &  \qquad (0 \leq q_1 \leq 1 \leq q_0 \leq2)
            \quad .
       \end{array}
       \right.
\end{equation}
The splitting function $\tilde{p}(q_0,q_1)$ is almost identical
to $p(q_0,q_1)$, their only difference being that the smaller 
splitting parameter $q_0$ is
now always drawn for the branching into the left subinterval whereas the
the larger splitting parameter $q_1=2-q_0$ goes into the right subinterval.
The one-point statistic is not affected by this modification of the
splitting function, since it does not care whether the larger splitting
parameter goes to the left or right subinterval. Hence the
multifractal exponents are not changed either. With (\ref{31fuenf}) we get
\begin{eqnarray}
\label{32fuenf}
  \tilde{\tau}(\nu)
    &=&  -\nu
         + \frac{1}{\ln 2}
         \ln\left(
         \int_0^2 {\rm d}q_0 {\rm d}q_1 \tilde{p}(q_0,q_1)
         (q_0^\nu + q_1^\nu)
         \right)
         \nonumber \\
    &=&  -\nu
         + \frac{1}{\ln 2}
         \ln\left(
         2 \int_0^1 {\rm d}q_0 p(q_0) (q_0^\nu + (2-q_0)^\nu)
         \right)
         \nonumber \\
    &=&  -(\nu -1)
         + \frac{1}{\ln 2}
         \ln\left(
         \int_0^2 {\rm d}q_0 p(q_0) q_0^\nu
         \right)
         \nonumber \\
    &=&  \tau(\nu)
         \quad .
\end{eqnarray}

We are hence forced to conclude that
a multifractal analysis does not suffice completely to characterize
the dynamics, i.e.\ the splitting function, of the underlying
multiplicative cascade process. 
This statement holds even more strongly once we also give up 
energy conservation in the splitting function, with the consequence
that properly defined backward moments show deviations from perfect
multifractal scaling. In order to attain the full information, there is
no way to get around spatial correlations in general, and the cumulant
correlation densities $C_{\bkappa_1, \ldots, \bkappa_n}$ in particular,
to extract the branching moments 
$\langle ( \ln q_0 )^{n-m} ( \ln q_1 )^{m} \rangle_c$
and hence the branching generating function
\begin{equation}
\label{32neun}
  Q[\lambda_0,\lambda_1]
    =  \sum_{n_1,n_2=0}^{\infty}
       \frac{1}{n_1!n_2!}
       \left\langle
       ( \ln q_0 )^{n_1} ( \ln q_1 )^{n_2}
       \right\rangle_c
       \lambda_0^{n_1} \lambda_1^{n_2}
       \quad .
\end{equation}
Using Eq.\ (\ref{221sechs}) the branching generating function can then
be uniquely inverted into the splitting function via a two-dimensional
inverse Laplace transformation,
\begin{equation}
\label{32zehn}
  \int_0^\infty dx \, dy \, p(2e^{-x}, 2e^{-y}) \,
  e^{-(\lambda_0 + 1)x - (\lambda_1 + 1)y }
    =  e^{ Q[\lambda_0, \lambda_1] - (\lambda_0 + \lambda_1 + 2) \ln 2 }
       \quad .
\end{equation}
As a consequence we view $Q[\lambda_0,\lambda_1]$ as the natural and
complete generalisation of the multifractal mass exponents $\tau(\nu)$
for random multiplicative binary cascade processes.

\section{Examples} 
\label{xmpl}

\subsection{Multiplicative binomial process: the $p$-model}

The $p$-model splitting function given in Eq.\ (\ref{21eins}),
when inserted into Eq.\ (\ref{221sechs}), determines the 
branching generating function 
$Q[\lambda_0,\lambda_1]$:
\begin{eqnarray}
\label{41eins}
  Q[\lambda_0,\lambda_1]
    &=&  \ln \left(
         \frac{1}{2}
         \left\{   
         \exp\left[
         \lambda_0 \ln (1+\beta)  +  \lambda_1 \ln (1-\beta)
         \right]
         \right.
         \right.
         \nonumber \\
    & &  \left.
         \left.
         \qquad
         + \exp\left[
         \lambda_0 \ln (1-\beta)  +  \lambda_1 \ln (1+\beta)
         \right]
         \right\}
         \rule{0mm}{1em}
         \right)
         \quad .
\end{eqnarray}
We introduce the transformation
\begin{equation}
\label{41zwei}
  \lambda_0  
    =  \frac{1}{2}
       \left(  \lambda_+ + \lambda_-  \right)
       \quad , \qquad
  \lambda_1
    =  \frac{1}{2}
       \left(  \lambda_+ - \lambda_-  \right)
       \quad ,
\end{equation}
which leads to
\begin{equation}
\label{41drei}
  Q[\lambda_0,\lambda_1]
    =  \frac{ \ln (1-\beta^2) }{ 2 }  \lambda_+
       + 
       \ln \left(
       \cosh \left(
       \frac{1}{2}  \lambda_-
       \ln\left( \frac{1+\beta}{1-\beta} \right)
       \right)
       \right)
       \quad .
\end{equation}
Fig.\ 2(a) illustrates this branching generating function.

\begin{figure}

\centerline{\epsfysize=150mm \epsfbox{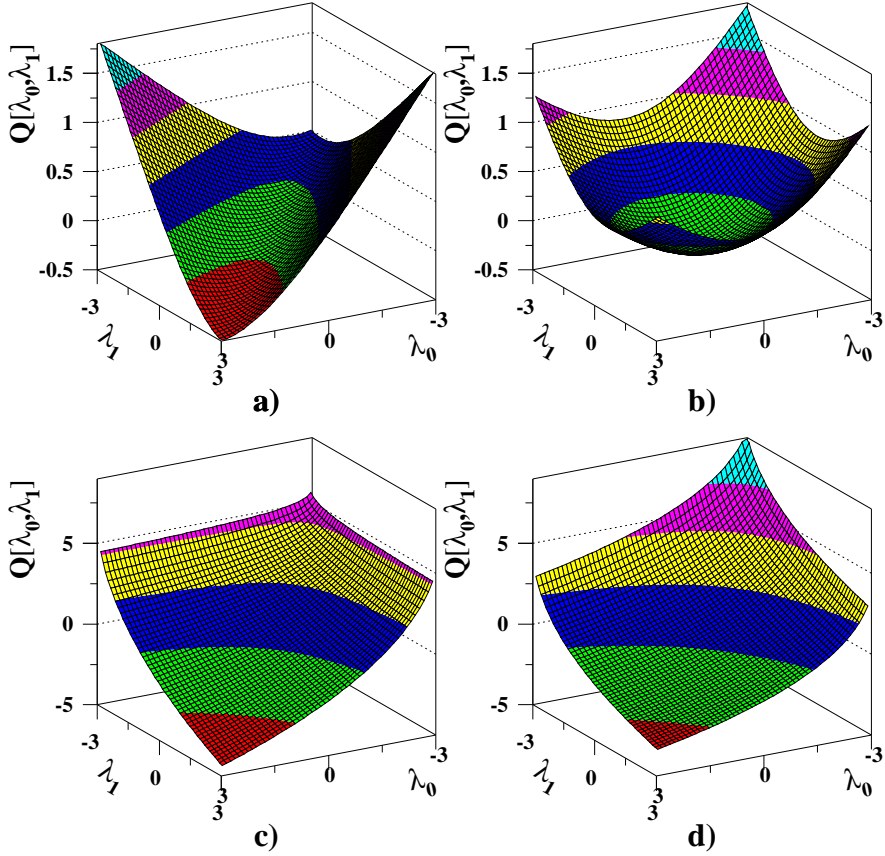}}

\caption{
Branching generating function $Q[\lambda_0,\lambda_1]$,
shown in the range $-3 \leq \lambda_0, \lambda_1 \leq 3$, for
(a) the $p$-model with $\beta=0.4$,
(b) the $\alpha$-model with $\beta=0.4$,
(c) the energy-conserving SRST cascade model with $\beta=3.2$, and
(d) the energy-non-conserving SRST cascade model with $\beta=3.2$.
}
\end{figure}

Now we calculate the derivatives of the branching generating function
with respect to the conjugate variables $\lambda_0$ and $\lambda_1$, 
which are needed for the cumulant densities 
$C_{\bkappa_1, \ldots, \bkappa_n}$. For the first derivative we find:
\begin{eqnarray}
\label{41vier}
  \left. 
  \frac{ \partial Q[\lambda_0,\lambda_1] }{ \partial \lambda_0 }
  \right|_{\lambda=0}
    &=&  \left. 
         \frac{ \partial Q[\lambda_0,\lambda_1] }{ \partial \lambda_1 }
         \right|_{\lambda=0}
         \nonumber \\
    &=&  \frac{ \ln (1-\beta^2) }{ 2 }
         + 
         \frac{1}{2}
         \ln\left( \frac{1+\beta}{1-\beta} \right) 
         \left.
         \tanh \left(
         \frac{1}{2} \lambda_-
         \ln\left( \frac{1+\beta}{1-\beta} \right) 
         \right)
         \right|_{\lambda=0}
         \nonumber \\
    &=&  \frac{1}{2}
         \ln(1 - \beta^2) 
         \quad .
\end{eqnarray}
For an arbitrary derivative of
$Q[\lambda_0,\lambda_1]$
of order $n_1$+$n_2 > 1$, we get the following relationship:
\begin{eqnarray}
\label{41fuenf}
  \left. 
  \frac{ \partial^{n_1+n_2} Q[\lambda_0,\lambda_1] }
       { \partial \lambda_0^{n_1} \partial \lambda_1^{n_2} }
  \right|_{\lambda=0}
    &=&  \left. 
         \frac{ \partial^{n_1+n_2} Q[\lambda_0,\lambda_1] }
              { \partial \lambda_-^{n_1+n_2} }
         \right|_{\lambda=0}
         \left(
         \frac{ \partial \lambda_- }{ \partial \lambda_0 }
         \right)^{n_1}
         \left(
         \frac{ \partial \lambda_- }{ \partial \lambda_1 }
         \right)^{n_2}
         \nonumber \\
    &=&  (-1)^{n_2}
         \left. 
         \frac{ \partial^{n_1+n_2} Q[\lambda_0,\lambda_1] }
              { \partial \lambda_0^{n_1+n_2} }
         \right|_{\lambda=0}
         \quad .
\end{eqnarray}
Except for an alternating sign, the two-point branching moments
$\langle (\ln q_0)^{n_1} (\ln q_1)^{n_2} \rangle_c$
are identical to the one-point branching moments
$\langle (\ln q_0)^{n_1+n_2} \rangle_c$;
this only holds for the $p$-model and not for the other three models
associated with the splitting functions (\ref{21zwei})--(\ref{21vier}).
As a consequence of (\ref{41fuenf}), we only need to calculate 
derivatives of
$Q[\lambda_0,\lambda_1]$
with respect to $\lambda_0$. We use the intermediate step of
(\ref{41vier}) and write for $n>1$:
\begin{equation}
\label{41sechs}
  \left. 
  \frac{ \partial^{n} Q[\lambda_0,\lambda_1] }{ \partial \lambda_0^{n} }
  \right|_{\lambda=0}
    =  \left.
       \frac{1}{2}
       \ln\left( \frac{1+\beta}{1-\beta} \right) \; 
       \left(
       \frac{ \partial^{n-1} }{ \partial \lambda_0^{n-1} }
       \left\{
       \tanh \left[
       \frac{1}{2} \lambda_0
       \ln\left( \frac{1+\beta}{1-\beta} \right) 
       \right]
       \right\}
       \right)
       \right|_{\lambda_0=0}
       \quad .
\end{equation}
Since $\tanh x$ is an odd function in $x$, i.e.\
\begin{eqnarray}
\label{41sieben}
  \tanh x
    & = &  x  -  \frac{x^3}{3}
           + \frac{2}{15} x^5
           - \frac{17}{315} x^7
           + \frac{62}{2835} x^9
           \mp \ldots
           \nonumber \\
    & = &  \sum_{m=1}^\infty
           \frac{ 2^{2m} (2^{2m}-1) B_{2m} }{ (2m)! }
           x^{2m-1}
\end{eqnarray}
with the Bernoulli numbers $B_{2m}$, all odd derivatives of the 
branching generating function
$Q[\lambda_0,\lambda_1]$
with respect to $\lambda_0$ vanish:

\begin{figure}

\centerline{\epsfysize=150mm \epsfbox{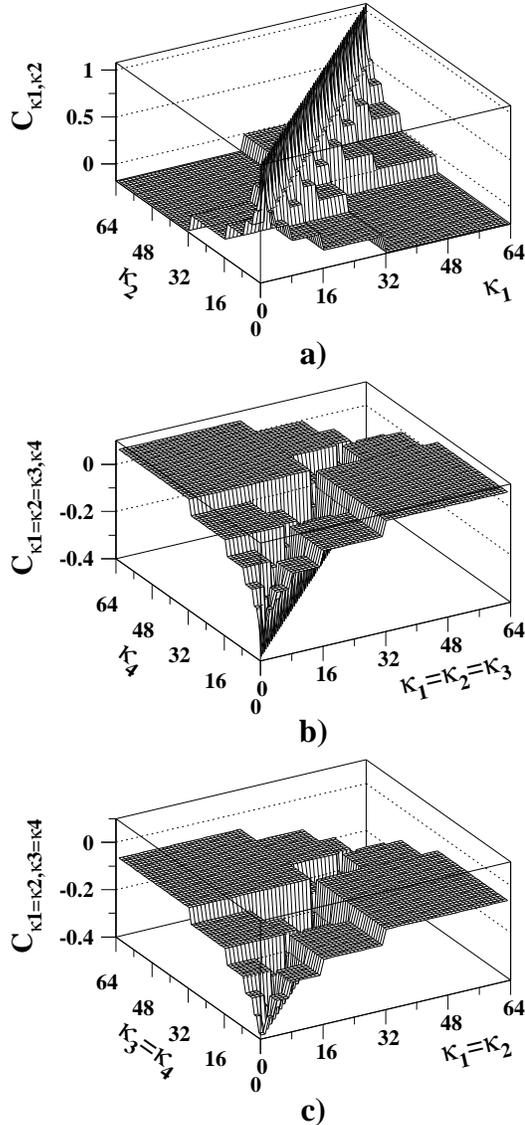}}

\caption{
Spatial cumulant densities
for the $p$-model with $\beta=0.4$ and $J=6$:
(a) second order $C_{\bkappa_1,\bkappa_2}$,
(b) fourth order with three equal indices, 
    $C_{\bkappa_1=\bkappa_2=\bkappa_3,\bkappa_4}$,
    and
(c) fourth order with two pairs of equal indices,
    $C_{\bkappa_1=\bkappa_2,\bkappa_3=\bkappa_4}$.
The indicated bin labels are related to the binary indices 
$\bkappa = (k_1 \cdots k_J)$
by $k = 1 + \sum_{j=1}^J k_j 2^{J-j}$.
}
\end{figure}

\begin{equation}
\label{41acht}
  \left. 
  \frac{ \partial^{n} Q[\lambda_0,\lambda_1] }{ \partial \lambda_0^{n} }
  \right|_{\lambda=0}
    =  0
       \qquad\qquad
       (n=3,5,7,\ldots)
       \quad .
\end{equation}
The same conclusion could have also been obtained directly from
(\ref{41fuenf}):
for odd $n \geq 3$ we have
$\partial^{n} Q[\lambda_0,\lambda_1] / 
 \partial \lambda_0^{n}|_{\lambda=0}
   =  - \partial^{n} Q[\lambda_0,\lambda_1] / 
      \partial \lambda_1^{n}|_{\lambda=0}$,
but from $q_0$/$q_1$-symmetry considerations of the splitting function
we expect
$\langle (\ln q_0)^{n} \rangle_c 
   =  \langle (\ln q_1)^{n} \rangle_c$;
hence,
$\partial^{n} Q[\lambda_0,\lambda_1] / 
 \partial \lambda_0^{n}|_{\lambda=0}
   =  \langle (\ln q_0)^{n} \rangle_c 
   =  0$. 
An immediate consequence of this result is that, for example, the
spatial cumulant densities of third order vanish,
$C_{\bkappa_1\bkappa_2\bkappa_3}=0$;
see Eq.\ (\ref{223sechs}).

For even derivatives of
$Q[\lambda_0,\lambda_1]$
with respect to $\lambda_0$ we find from (\ref{41sechs}) and 
(\ref{41sieben}) that
\begin{equation}
\label{41neun}
  \left. 
  \frac{ \partial^{n} Q[\lambda_0,\lambda_1] }{ \partial \lambda_0^{n} }
  \right|_{\lambda=0}
    =  \frac{(2^n-1)B_n}{n}
       \left(
       \ln \left( \frac{1+\beta}{1-\beta} \right) 
       \right)^n
       \qquad
       (n=2,4,6,\ldots)
       \quad .
\end{equation}

The results (\ref{41vier}), (\ref{41fuenf}), (\ref{41acht}) and 
(\ref{41neun}) can now be inserted into Eqs.\ 
(\ref{223vier})--(\ref{223sieben}) to determine the cumulant correlation
densities
$C_{\bkappa_1, \ldots, \bkappa_n}
 = \langle (\ln\epsilon_{\bkappa_1}) \cdots (\ln\epsilon_{\bkappa_n})
   \rangle_c$.
Fig.\ 3 illustrates the results for the two-point statistics:
$C_{\bkappa_1,\bkappa_2}$ of second order, and 
$C_{\bkappa_1=\bkappa_2=\bkappa_3,\bkappa_4}$ and
$C_{\bkappa_1=\bkappa_2,\bkappa_3=\bkappa_4}$
of fourth order; note again that the third order
$C_{\bkappa_1,\bkappa_2,\bkappa_3}$,
vanishes. Figs.\ 4(a) and 4(b) show the second-order
and fourth order cumulants as a function of the ultrametric distances 
respectively.

\begin{figure}

\centerline{\epsfysize=150mm \epsfbox{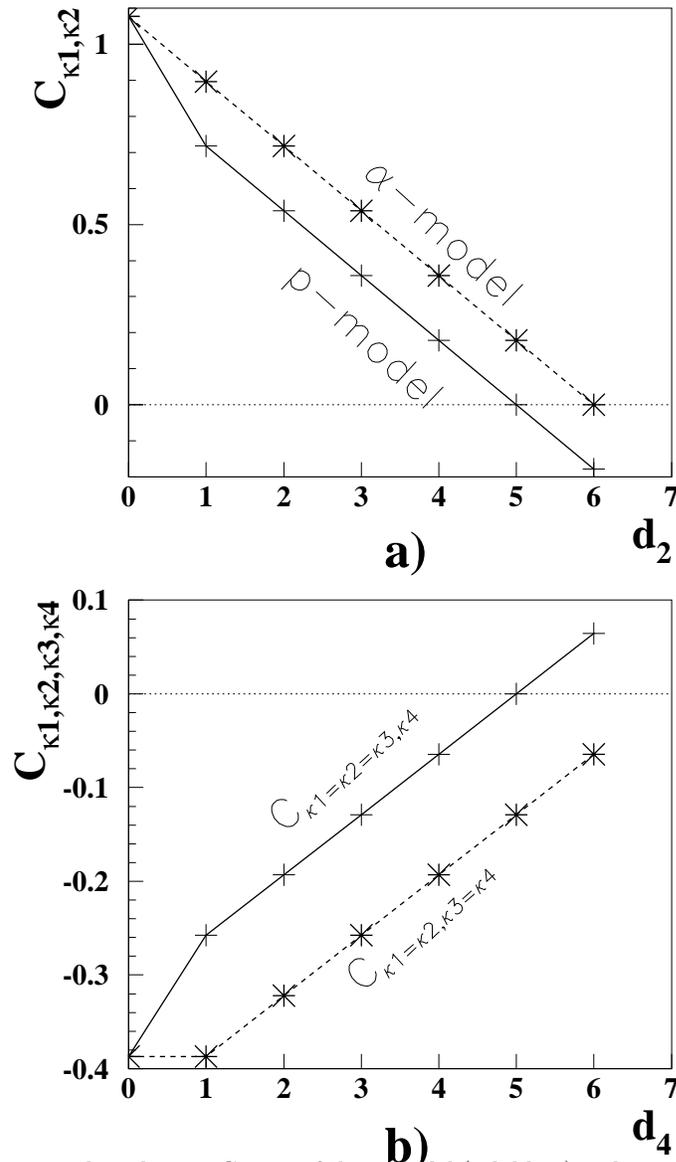}}

\caption{
(a) Second-order cumulant density $C_{\bkappa_1, \bkappa_2}$ 
of the $p$-model (solid line) and 
$\alpha$-model (dashed line)
as a function of the ultrametric distance 
$d_2 = {\rm dist}(\bkappa_1,\bkappa_2)$.
Parameters values $\beta=0.4$ and $J=6$ were used.
(b) Two projections of fourth-order cumulant densities
for the $p$-model as a function of the ultrametric distance $d_4$.
}
\end{figure}

Since the $p$-model splitting function is energy-conserving and 
$q_0/q_1$-symmetric we can use relation (\ref{32zwei}) to extract the
multifractal exponents $\tau(\nu)$ from the branching generating
function $Q[\lambda_0,\lambda_1]$. This leads to the well-known result
\cite{FED88,MEN87}
\begin{equation}
\label{32acht}
  \tau(\nu)
    =  \frac{1}{\ln 2}
       \ln\left[
       \left( \frac{1+\beta}{2} \right)^\nu
       +
       \left( \frac{1-\beta}{2} \right)^\nu
       \right]
       \quad .
\end{equation}
It is illustrated in Fig.\ 5. Via 
\begin{eqnarray}
\label{32drei}
  \left. \frac{\partial^n \tau (\nu)}{\partial \nu^n} \right|_{\nu=0}
    & = &  - \delta_{n,1}
           + 
           \frac{1}{\ln 2}  
           \left.
           \frac{ \partial^n Q[\lambda_0,\lambda_1] }
                { \partial \lambda_0^n }
           \right|_{\lambda=0}
           \nonumber  \\
    & = &  - \delta_{n,1}
           + 
           \frac{1}{\ln 2}
           \left\langle ( \ln q_0 )^n \right\rangle_c
\end{eqnarray}
the derivatives
of $\tau(\nu)$ with respect to $\nu$ are linked to the derivatives
(\ref{41vier}), (\ref{41acht}) and (\ref{41neun}) of the $p$-model
branching generating function (\ref{41drei}).

\begin{figure}

\centerline{\epsfysize=150mm \epsfbox{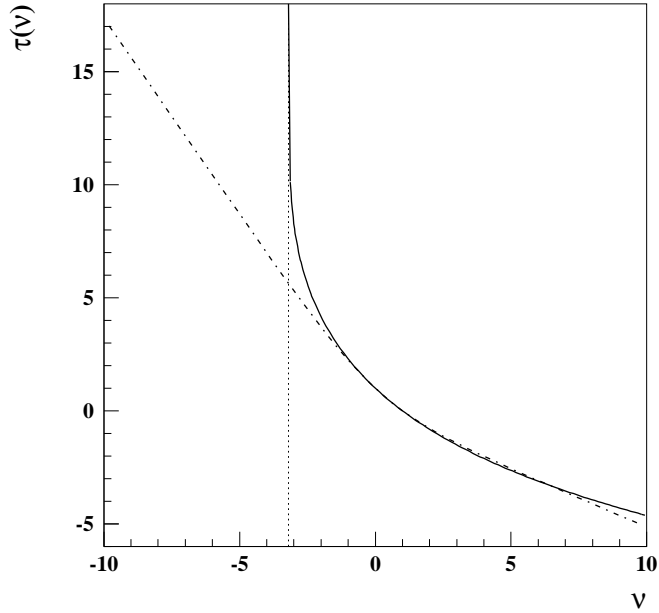}}

\caption{
Multifractal exponents $\tau(\nu)$ of the $p$-model (dash-dotted line)
with $\beta=0.4$ and of the SRST cascade model (full line) with
$\beta=3.2$.
}
\end{figure}

\subsection{$\alpha$-model}

The symmetric $\alpha$-model is similar to the $p$-model
except that it does not conserve energy in a cascade
splitting. From the splitting function (\ref{21zwei}), its
b.g.f.\ is found to be
\begin{eqnarray}
\label{42eins}
  Q[\lambda_0,\lambda_1]
    &=&  {\textstyle{1\over 2}} (\lambda_0+\lambda_1) \ln(1-\beta^2)
         \nonumber \\
    & &  +
         \ln \left\{
         \cosh\left[ {1\over 2} \lambda_0
         \ln\left( 1 + \beta \over 1 - \beta \right)
         \right]
         \right\}
         \nonumber \\
    & &  +
         \ln \left\{
         \cosh\left[ {1\over 2} \lambda_1
         \ln\left( 1 + \beta \over 1 - \beta \right)
         \right]
         \right\}
         \quad ,
\end{eqnarray}
clearly different from the b.g.f.\ (\ref{41drei})
for the $p$-model. Fig.\ 2(b) depicts the b.g.f.\ (\ref{42eins}).
Note that for $\lambda_0 = 0$
or $\lambda_1 = 0$ the two expressions (\ref{41drei})
and (\ref{42eins}) become identical; consequently,
the one-point branching moments
of the $\alpha$-model are identical to those of the $p$-model
given in (\ref{41vier}), (\ref{41acht}) and (\ref{41neun}):
\begin{equation}
\label{42zwei}
  \left(
  \left.
  { \partial^n Q[\lambda_0, \lambda_1] \over
    \partial \lambda_0^n}
  \right|_{\lambda = 0}
  \right)_{\alpha{\rm -model}}
    =  \left(
       \left.
       { \partial^n Q[\lambda_0, \lambda_1] \over
         \partial \lambda_0^n}
       \right|_{\lambda = 0}
       \right)_{p{\rm -model}}
       \,.
\end{equation}
This is the
reason why, in a multifractal approach,  the two models look the same
asymptotically. To see differences between them,
one must consider two-point branching moments: for the 
$\alpha$-model they all vanish,
\begin{equation}
\label{42drei}
  \left.
  { \partial^n Q[\lambda_0, \lambda_1] \over
    \partial \lambda_0^{n_1} \partial \lambda_1^{n - n_1}}
  \right|_{\lambda = 0}
    =  \left\langle (\ln q_0)^{n_1} (\ln q_1)^{n - n_1} \right\rangle_c
    =  0
       \quad ,
\end{equation}
where $1 \leq n_1 < n$. We hence see that the two-point
branching moments are sensitive to the violation of energy
conservation in the splitting function.
As a consequence of (\ref{42zwei}) and (\ref{42drei}), the cumulant
densities $C_{\bkappa_1, \ldots, \bkappa_n}$ of the $\alpha$- and $p$-model
now look slightly different. Fig.\ 4(a) compares $C_{\bkappa_1, \bkappa_2}$
of second order as a function of the ultrametric distance (\ref{223eins}).

\subsection{SRST cascade}
\label{srstec}

The $p$-model is able to
describe the multifractal aspects of the intermittent fluctuations 
occurring in the energy dissipation field in fully developed turbulence. 
However, due to its simplicity, the experimental multiplier distributions
cannot be reproduced. In Ref.\ \cite{SRE95}, a modification of the
$p$-model was proposed which accounts for the correct multiplier
distributions; we call this modification the SRST cascade model.

Insertion of its splitting function (\ref{21drei}) into 
Eq.\ (\ref{221sechs}) yields an 
analytic expression for its b.g.f.,
\begin{equation}
\label{43eins}
  Q[\lambda_0,\lambda_1]
    =  \ln\left[
       \int_0^1 {\rm d}z 
       z^{\lambda_0+\beta-1}  (1-z)^{\lambda_1+\beta-1}
       \right]
       +
       \ln\left[
       \frac{\Gamma(2\beta)}{\Gamma(\beta)^2}
       \right]
       +
       ( \lambda_0 + \lambda_1 ) \ln 2
       \quad .
\end{equation}
Since $\beta=3.2$ and $\lambda_0 \approx 0 \approx \lambda_1$, so that
$\lambda_0+\beta > 0$ and $\lambda_1+\beta > 0$,
the integral appearing in the first term of the right hand side can be
identified with the beta function
$B(\lambda_0+\beta,\lambda_1+\beta)
 = \Gamma(\lambda_0+\beta)
   \Gamma(\lambda_1+\beta)
   / \Gamma(\lambda_0+\lambda_1+2\beta)$.
This leads to
\begin{equation}
\label{43zwei}
  Q[\lambda_0,\lambda_1]
    =  ( \lambda_0 + \lambda_1 ) \ln 2
       +
       \ln\left(
       \frac{\Gamma(2\beta)}{\Gamma(\lambda_0+\lambda_1+2\beta)}
       \frac{\Gamma(\lambda_0+\beta)}{\Gamma(\beta)}
       \frac{\Gamma(\lambda_1+\beta)}{\Gamma(\beta)}
       \right)
       \quad .
\end{equation}
This result is illustrated in Fig.\ 2(c).

{}From (\ref{43zwei}) the spatial cumulant densities
$C_{\bkappa_1, \ldots, \bkappa_n}$
can be calculated in the straightforward manner presented in
Sec.~\ref{cumcord}.
In  the lowest even orders, the results are very similar to those
obtained for the $p$-model; the odd orders, however, are not equal to 
zero anymore.

Making use of the relationship (\ref{32zwei}) the result (\ref{43zwei}) 
translates into the following multifractal exponents:
\begin{equation}
\label{43drei}
  \tau(\nu)
    =  1 
       + \frac{1}{\ln 2} 
       \ln\left(
       \frac{\Gamma(2\beta)}{\Gamma(\nu+2\beta)}
       \frac{\Gamma(\nu+\beta)}{\Gamma(\beta)}
       \right)
       \quad .
\end{equation}
Note that this expression, which is illustrated in 
Fig.\ 5, is only defined for $\nu + \beta > 0$; for $\nu + \beta \leq 0$
the integral in (\ref{43eins}) diverges.
This is equivalent to the statement that the negative moments
with $\nu \leq -\beta$ 
of the splitting functions (\ref{21drei}) and (\ref{21vier})
do not exist.
For the multifractal spectrum to exist over the full 
$\nu$ range, however, moments of all orders, both positive and negative,
must be finite. The absence of finite negative moments hence implies that
it is not possible to construct full $\tau(\nu)$ and
$f(\alpha)$ curves for this specific splitting function.
It is thus an example of a well-defined selfsimilar cascade
process which cannot be described fully by the multifractal formalism.

\subsection{SRST cascade with no energy conservation}

Experimentally, the intermittent structures in the three-dimensional 
energy dissipation
field of fully developed turbulence are observed on a one-dimensional
cut. Although energy is conserved in three dimensions, this is probably not
the case in one dimension. For the multiplicative branching models, this
has the consequence that the splitting function $p(q_0,q_1)$ cannot
be expected to conserve energy. In this spirit, the expression (\ref{21vier})
represents an untested extrapolation of the SRST multiplier distribution
(\ref{21drei}), which has been deduced from one-dimensional data under
the assumption of energy conservation \cite{SRE95}.

Insertion of the splitting function (\ref{21vier}) into Eq.\ 
(\ref{221sechs}) yields the corresponding branching generating function:
\begin{equation}
\label{44eins}
  Q[\lambda_0,\lambda_1]
    =  ( \lambda_0 + \lambda_1 ) \ln 2
       +
       \ln\left(
       \frac{\Gamma(2\beta)}{\Gamma(\lambda_0+2\beta)}
       \frac{\Gamma(\lambda_0+\beta)}{\Gamma(\beta)}
       \right)
       +
       \ln\left(
       \frac{\Gamma(2\beta)}{\Gamma(\lambda_1+2\beta)}
       \frac{\Gamma(\lambda_1+\beta)}{\Gamma(\beta)}
       \right)
       \quad ;
\end{equation}
the explicit derivation is analogous to the one given in 
Sec.~\ref{srstec}.
The illustration of (\ref{44eins}) is given in Fig.\ 2(d); it differs
from the branching generating function (\ref{43zwei}) of the 
SRST cascade model with energy conservation. As in the $p$/$\alpha$-model
comparison, the one-point derivatives
\begin{equation}
\label{44zwei}
  \left(
  \left.
  { \partial^n Q[\lambda_0, \lambda_1] \over
    \partial \lambda_0^n}
  \right|_{\lambda = 0}
  \right)_{\rm SRST \atop (no EC)}
    =  \left(
       \left.
       { \partial^n Q[\lambda_0, \lambda_1] \over
         \partial \lambda_0^n}
       \right|_{\lambda = 0}
       \right)_{\rm SRST \atop (EC)}
\end{equation}
of the SRST cascade model with (EC) and without (no EC) energy conservation
are identical, while for the two-point derivatives we find
($1 \leq n_1 < n$)
\begin{equation}
\label{44drei}
  \left(
  \left.
  { \partial^n Q[\lambda_0, \lambda_1] \over
    \partial \lambda_0^{n_1}
    \partial \lambda_1^{n-n_1}}
  \right|_{\lambda = 0}
  \right)_{\rm SRST \atop (no EC)}
    =  0
    \neq  
       \left(
       \left.
       { \partial^n Q[\lambda_0, \lambda_1] \over
         \partial \lambda_0^{n_1}
         \partial \lambda_1^{n-n_1}}
       \right|_{\lambda = 0}
       \right)_{\rm SRST \atop (EC)}
       \quad .
\end{equation}
Consequently, the cumulant densities
$C_{\bkappa_1, \ldots, \bkappa_n}$
of the SRST-cascade model with and without energy conservation are
different.

\section{Conclusions}     
\label{concll}

With a clever change of variables from energy densities
$\epsilon_{\bkappa}$ to the singularity strengths
$\alpha_{\bkappa}$ or $\ln\epsilon_{\bkappa}$, we have derived an
analytic expression for the multivariate generating function of
binary multiplicative cascade models. The latter completely describes
the $n$-point statistics, i.e.\ the spatial (cumulant) correlation
densities of arbitrary order $n$. The key input has been a bivariate
branching generating function, which is related to the underlying
splitting function of the binary multiplicative cascade process
via a two-dimensional Laplace transform. This branching
generating function can be understood as a natural and, for 
selfsimilar binary cascade processes, complete generalisation of the
multifractal mass exponents. While its properties completely fix the
spatial correlation densities, the multifractal mass exponents do not. 
Various cascade models, relevant to fully developed turbulence, 
have been discussed to underpin this point.

We have shown that, given that the experimentally
measurable cumulants in $(\ln \epsilon_{k_1 \cdots k_J}^{(J)})$
are $n$-fold derivatives of the branching generating function, 
the latter can in principle be reconstructed from the former.
With the help of Eqs.\ (\ref{32neun}) and (\ref{32zehn}), 
the b.g.f.\ can then 
be inverted into the splitting function via a two-dimensional 
inverse Laplace transform. 
In this way, the violation of energy conservation along 
one-dimensional cuts through the three-dimensional energy dissipation
field can be inspected.

Before this new approach is applied directly to (turbulence) data,
however, a number of complications will have to be dealt with. In order 
to infer the branching generating function from the cumulant densities
$C_{\bkappa_1, \ldots, \bkappa_n}$ (to all orders, in principle),
a very effective representation of the latter has to be found; here,
as in Refs.\ \cite{GRE95,GRE96}, a wavelet transformation might be
useful to compress the information contained in the cumulant
densities.

Moreover, there is the problem of non-homogeneity:
as a consequence of the hierarchical nature of the cascade evolution, 
the theoretical correlation functions are not invariant with respect to 
spatial translations, in contradiction to experimental measurements.
It remains to be seen whether and in what way a scheme to restore 
homogeneity, as for example the one used in \cite{GRE97a}, influences 
or destroys the capability of inferring the branching generating function.

Furthermore, the observed statistical dependence of multipliers
\cite{SRE95,PED96,NEL96} has to be taken into account. Recent
simulations \cite{JOU98} indicate that this issue is closely linked
to the restoration of homogeneity. This is in agreement with the conclusions
reached by Nelkin and Stolovitzky \cite{NEL96} by a different route,
who argue that the experimentally proven dependence of multiplier 
distributions on the position of the subinterval implies that the 
multipliers are not statistically independent. 
Since any scheme to restore homogeneity will necessarily average out 
subinterval positions in some way, it will likely influence the multipliers'
statistical dependence also. This remains to be explored in detail
\cite{JOU98}.

Finally, the binary
structure of the selfsimilar cascade processes discussed in this 
paper may not be appropriate: assuming that the physical processes 
themselves are, indeed, selfsimilar, the best selfsimilar basis for a 
scaling analysis (such as a specific wavelet) should be 
selected by the data itself.

Once these points are clarified, new information can hopefully
be gleaned from the analysis of ``fully developed turbulence
data''. Besides fully-developed turbulence,
we envisage many and diverse applications of our analytic solution 
in other branches of physics. The case of QCD branching processes 
immediately comes to mind. For the latter, 
the $\alpha$- and $p$-models have already been used in this
context as simulation toy models \cite{BIA86,LIP89}.
Implications in this and, for example, random multiplicative 
process calculations in large-scale structure formation in the universe
\cite{PAN97} remain to be explored.

\acknowledgements
This work was supported in part by the South African Foundation for
Research Development. PL acknowledges support by APART of the Austrian
Academy of Sciences.

\newpage


\begin{thebibliography}{99}


\bibitem{FED88}
  J.\ Feder, {\em Fractals} (Plenum Press, New York, 1988).

\bibitem{MEN91}C.\ Meneveau and K.R.\ Sreenivasan,
         J.\ Fluid Mech.\ {\bf 224}, 429 (1991). 

\bibitem{FRI95}
  U.\ Frisch, {\em Turbulence} 
  (Cambridge University Press, Cambridge, 1995).

\bibitem{CAT87}
  M.E.\ Cates and J.M.\ Deutsch,
  Phys.\ Rev.\ A{\bf 35}, 4907 (1987).
  
\bibitem{MEN90}
  C.\ Meneveau and A.B.\ Chhabra,
  Physica A {\bf 164}, 564 (1990).

\bibitem{NEI93}
  J.\ O'Neil and C.\ Meneveau,
  Phys.\ Fluids A {\bf 5}, 158 (1993).

\bibitem{FEI87}M.J.\ Feigenbaum,
         J.\ Stat.\ Phys.\ {\bf 46}, 919 (1987); 
                           {\bf 46}, 925 (1987);
         A.B.\ Chhabra, R.V.\ Jensen and K.R.\ Sreenivasan,
         Phys.\ Rev. A{\bf 40}, 4593 (1989). 

\bibitem{MEN87}C.\ Meneveau and K.R.\ Sreenivasan,
         Phys.\ Rev.\ Lett.\ {\bf 59}, 1424 (1987).

\bibitem{SCH85}D.\ Schertzer and S.\ Lovejoy, 
        in {\it Turbulent Shear Flows 4}, 1985, edited by
        L.J.S.\ Bradbury, F.\ Durst, B.\ Launder, F.W.\ Schmidt
        and J.H.\ Whitelaw, (Springer, 1985), pp.\ 7--33.

\bibitem{GRE95}M.\ Greiner, P.\ Lipa and P.\ Carruthers,
        Phys.\ Rev.\ E {\bf 51}, 1948 (1995).

\bibitem{GRE96}M.\ Greiner, J.\ Giesemann, P.\ Lipa and P.\ Carruthers,
        Z.\ Phys.\ C {\bf 69}, 305 (1996).

\bibitem{GRE97b}
  M.\ Greiner, H.C.\ Eggers, and P.\ Lipa, 
  preprint mpi-pks/9712007,  STPHY 27/97, HEPHY-PUB 676/97.

\bibitem{SRE95}
  K.R.\ Sreenivasan and G.\ Stolovitzky,
  J.\ Stat.\ Phys.\ {\bf 78}, 311 (1995).

\bibitem{GIE97}
  J.~Giesemann, M.~Greiner and P.~Lipa,
  Physica A {\bf 247}, 41 (1997) .

\bibitem{NOV71}E.A.\ Novikov,
        Prikl.\ Mat.\ Mekh.\ {\bf 35}, 266 (1971);
        Phys.\ Fluids A{\bf 2}, 814 (1990).

\bibitem{GRE97a}
  M.\ Greiner, J.\ Giesemann, and P.\ Lipa, 
  Phys.\ Rev.\ E {\bf 56}, 4263 (1997).

\bibitem{PED96}G.\ Pedrizzetti, E.A.\ Novikov and A.A.\ Praskovsky,
        Phys.\ Rev.\ E{\bf 53}, 475 (1996).

\bibitem{NEL96}M.\ Nelkin and G.\ Stolovitzky,
        Phys.\ Rev.\ E{\bf 55}, 5100 (1996).

\bibitem{JOU98}B.\ Jouault, M.\ Greiner and P.\ Lipa,
        (in preparation).

\bibitem{BIA86}
  A.\ Bia\l as and R.\ Peschanski,
  Nucl.~Phys.\ B {\bf 273}, 703 (1986);
  {\bf 308}, 857 (1988).

\bibitem{LIP89}P.\ Lipa and B.\ Buschbeck,
        Phys.\ Lett.\ B {\bf 223}, 465 (1989).

\bibitem{PAN97}J.\ Pando, P.\ Lipa, M.\ Greiner and L.-Z.\ Fang,
        Ap.\ J.\ (in press).


\end{thebibliography}
\end{document}